\documentclass[intlimits,twoside,a4paper]{article}

\usepackage{amsmath,amssymb,amsbsy}
\usepackage{graphicx}
\usepackage{wrapfig}
\usepackage{color}
\usepackage{epstopdf}
\usepackage[T2A]{fontenc}
\usepackage[cp1251]{inputenc}

\usepackage[eqsecnum]{cmpj2}

\issue{2015}{18}{3}{33701}
\doinumber{10.5488/CMP.18.33701}

\newcommand{\dxy}{\ensuremath{d_{xy}}}
\newcommand{\dxz}{\ensuremath{d_{xz}}}
\newcommand{\dyz}{\ensuremath{d_{yz}}}
\newcommand{\dzz}{\ensuremath{d_{3z^2-1}}}
\newcommand{\dxxyy}{\ensuremath{d_{x^2-y^2}}}

\newcommand{\mb}{\ensuremath{\mu_{\text{B}}}}

\title[La$_{1-x}$Pr$_x$Co$_2$P$_2$ phosphides]{Electronic structure, magnetic ordering and X-ray magnetic
  circular dichroism in La$_{1-x}$Pr$_x$Co$_2$P$_2$ phosphides}

  \author{L.V. Bekenov, S.V. Moklyak, V.N. Antonov}

\address{ Institute of Metal Physics of the National
Academy of Sciences of Ukraine, 36 Vernadsky Blvd., 03142 Kyiv, Ukraine}

\authorcopyright{L.V. Bekenov, S.V. Moklyak, V.N. Antonov, 2015}
\date{Received January 31, 2015}

\begin{document}

\maketitle

\begin{abstract}

The electronic structure and magnetic ordering in
La$_{1-x}$Pr$_x$Co$_2$P$_2$ ($x=0, 0.25$, and 1) phosphides have been
studied theoretically using the fully relativistic spin-polarized
Dirac linear muffin-tin orbital (LMTO) band-structure method. The
X-ray absorption and X-ray magnetic circular dichroism spectra at the
Co$L_{2,3}$ and Pr$M_{4,5}$ edges have been investigated theoretically within the
framework of the LSDA+$U$ method. The core-hole effect in the final state
as well as the effects of the electric quadrupole $E_2$ and magnetic
dipole $M_1$ transitions have been investigated. Good agreement with
experimental measurements has been found.

\pacs 71.28.+d, 71.25.Pi, 75.30.Mb

\end{abstract}

\section{Introduction}

\label{sec:introd}

Ternary intermetallics AT$_2$X$_2$, where A $-$ alkali, alkali-earth,
rare-earth, or actinide metal, T $-$ transition metal, and X $-$ nonmetal,
often demonstrate intrinsically complex magnetic structures and a wide
variety of physical properties. They belong to the ThCr$_2$Si$_2$
structure type and can be conveniently represented as materials built
by stacking covalently bonded transition metal-metalloid T$_2$X$_2$
layers, made of edge-sharing TX$_4$ tetrahedra, with ionic A
atoms. This body-centered tetragonal structure is rather simple with
only one crystallographic site for each atomic species and only one
variable positional parameter, the $z$ value of the X site. Due to
the great variety of possible elements in this structure, over 1000 different compounds of this type are known
already \cite{book:VilCal85}. Some of these compounds exhibit
fascinating physical properties such as superconductivity, valence
fluctuations, local and itinerant magnetism. Interest to these
materials has been reinvigorated recently due to the discovery of
non-Fermi-liquid behavior in YbRh$_2$Si$_2$ \cite{GGW+03} and
high-temperature superconductivity in K-doped BaFe$_2$As$_2$
\cite{RLY+08,RTJ08}. In addition, materials with layered arrangements
of magnetic moments are also of great interest because they tend
to exhibit a highly anisotropic magnetic behavior. The magnetic
properties of the silicides and germanides possessing this structure were
studied by many groups
\cite{book:SzLe84,book:PaCh84,FOS+85,DDS89,MAI+97,WYC+97,DAD+07}.
Besides superconductivity, AT$_2$X$_2$ phases exhibit a plethora of
other phase transition phenomena. Thus, SmMn$_2$Ge$_2$ becomes
ferromagnetic below 348 K and then undergoes a transition to an
antiferromagnetic state at 196 K, followed by a re-entrant
ferromagnetic transition at 64 K \cite{FOS+85}. This peculiar magnetic
behavior stems from the layered structure and the presence of two
magnetic sublattices in these materials and was later observed for a
number of rare-earth manganese germanides and silicides \cite{DDS89,MAI+97,WYC+97,DAD+07}.
Such sequential magnetic transitions
have not been observed in isostructural phosphides which might
explain why their magnetic properties have received somewhat less attention
than the tetrelides. The isotypic phosphides were prepared
much later \cite{RJM+92} and the investigations of their magnetic
properties have started only recently.

In the compounds RT$_2$P$_2$ (R = lanthanides, T = Fe, Co, Ni) with the
trivalent rare earth elements, magnetic moments at the transition metal sites
were observed only for the cobalt containing compounds
\cite{JeRe87,MMM+88,ReJe90}. Reehuis  et al. carried out a comprehensive
study of the structural and magnetic properties of ternary RCo$_2$P$_2$
materials \cite{ReJe90,RBJ+93,EEB+94}. Recently, Kovnir  et al.
\cite{KTZ+10,KRM+11,KGT+11} and other groups \cite{JCL+10,JJS+11} have
demonstrated that the magnetic properties of the ThCr$_2$Si$_2$ type
phosphides can be as rich and diverse as those of the aforementioned silicides
and germanides, provided that proper iso- and aliovalent substitutions are
made to modify the crystal and electronic structures of the materials.
Ternary cobalt phosphides RCo$_2$P$_2$ appear to be on the verge of magnetic
instability \cite{KGT+11}. Indeed, LaCo$_2$P$_2$ is characterized by
ferromagnetic ordering of Co magnetic moments at 132 K \cite{KTZ+10,MMM+88},
while other representatives of this family (R = Ce, Pr, Nd, Sm) order
antiferromagnetically at around room temperature \cite{ReJe90} (with the
exception of the Ce-containing compound, which shows an antiferromagnetic
transition at 440 K purportedly due to the mixed valence of Ce). A
substantial difference is also observed in the crystal structures of these
compounds \cite{JMM+85}. In LaCo$_2$P$_2$, the [Co$_2$P$_2$] layers are far
from each other, with the interlayer P--P distance of 3.16~\AA\, indicating
essentially no bonding between the phosphorus atoms. By contrast, other
RCo$_2$P$_2$ structures show a weakly covalent P--P interaction at relatively
small interplanar P--P distances (2.57~\AA\, in PrCo$_2$P$_2$
\cite{KTZ+10}). Thus, a drastic change in the magnetic behavior is
accompanied by the formation of the P--P bonding pathway between the
[Co$_2$P$_2$] layers.

Kovnir  et al. \cite{KTZ+10} embarked on the study of structural,
magnetic, and electronic properties of quaternary compositions
La$_{1-x}$Pr$_x$Co$_2$P$_2$ ($0 \leqslant x \leqslant 1$). The choice of Pr was based on
the fact that it does not exhibit a mixed valence like Ce, and, therefore, is the
closest rare-earth neighbor of La in both the size and the formal ionic
charge. They focused on the drastically different magnetic properties of
LaCo$_2$P$_2$ and PrCo$_2$P$_2$, which exhibit ferromagnetic ($T_{\mathrm{C}}$ = 132 K)
and antiferromagnetic ($T_{\mathrm{N}}$ = 305 K) ordering in the Co sublattice,
respectively \cite{KTZ+10}. In both cases, a ferromagnetic arrangement of
magnetic moments within the square plane of Co atoms is observed. In
LaCo$_2$P$_2$, the Co moments are aligned in-plane and parallel to the moments
in the other layers, leading to the ferromagnetic ground state \cite{EEB+94}.
In PrCo$_2$P$_2$, they are aligned perpendicular to the plane (along the $c$
axis), but the antiparallel magnetic coupling between the neighboring planes results
in antiferromagnetism \cite{RBJ+93}. In \cite{KTG+13},
magnetic behavior of La$_{0.75}$Pr$_{0.25}$Co$_2$P$_2$ was
investigated by a combination of magnetic measurements, magneto-optical
imaging, neutron diffraction, and X-ray absorption spectroscopy, including
X-ray magnetic circular dichroism.

The aim of this paper is a theoretical study, from the first principles, of
the electronic structure, magnetic ordering and X-ray magnetic circular
dichroism in La$_{1-x}$Pr$_x$Co$_2$P$_2$ ($x=0, 0.25$, and 1)
compounds.  The energy band structure of La$_{1-x}$Pr$_x$Co$_2$P$_2$
($x=0, 0.25$, and 1) compounds is calculated within the {\it ab initio} approach
taking into account strong electron correlations by applying the local spin-density
approximation (LSDA) to the density functional theory supplemented by a Hubbard $U$ term
(LSDA+$U$) \cite{AZA91}. The paper is organized as follows. The computational
details are presented in section~2. Section~3 presents the electronic structure,
XAS and XMCD spectra of La$_{1-x}$Pr$_x$Co$_2$P$_2$ ($x=0, 0.25$, and 1)
compounds calculated in the LSDA+$U$ approximation. The theoretical results are
compared to experimental measurements. Finally, the results are summarized
in section~4.

\section{Computational details}
\label{sec:details}

\paragraph{X-ray magnetic circular dichroism.}

The absorption coefficient $\mu^{\lambda}_j (\omega)$ for incident X-ray of
polarization $\lambda$ and photon energy $\hbar \omega$ can be determined as
the probability of electronic transitions from initial core states with the
total angular momentum $j$ to final unoccupied Bloch states:
%\begin{equation}
\begin{eqnarray}
\mu^j_{\lambda} (\omega) &=& \sum_{m_j} \sum_{n \bf k} \left| \left\langle \Psi_{n \bf k} |
\Pi _{\lambda} | \Psi_{jm_j} \right\rangle \right|^2 \delta (E _{n \bf k} - E_{jm_j} -
\hbar \omega )
%\nonumber \\ &&\times
\theta (E _{n \bf k} - E_{\mathrm{F}} ) \, ,
\label{mu}
\end{eqnarray}
%\end{equation}
where $\Psi _{jm_j}$ and $E _{jm_j}$ are the wave function and the
energy of a core state with the projection of the total angular
momentum $m_j$; $\Psi_{n\bf k}$ and $E _{n \bf k}$ are the wave
function and the energy of a valence state in the $n$-th band with the
wave vector $\bf k$; $E_{\mathrm{F}}$ is the Fermi energy.
$\Pi _{\lambda}$ is the electron-photon interaction
operator in the dipole approximation
\begin{equation}
\Pi _{\lambda} = -\re {\boldsymbol\alpha} \bf {a_{\lambda}}\,,
\label{Pi}
\end{equation}
where $\boldsymbol{\alpha}$ are the Dirac matrices, $\bf {a_{\lambda}}$ is the
$\lambda$ polarization unit vector of the photon vector potential,
with $a_{\pm} = 1/\sqrt{2} (1, \pm i, 0),
a_{\parallel}=(0,0,1)$. Here, $+$ and $-$ denotes, respectively, left
and right circular photon polarizations with respect to the
magnetization direction in the solid. The X-ray magnetic circular
and linear dichroism are given by $\mu_{+}-\mu_{-}$ and
$\mu_{\parallel}-(\mu_{+}+\mu_{-})/2$, respectively.  Detailed
expressions of the matrix elements in the electric dipole approximation
may be found in \cite{GET+94,ABP+93,book:AHY04,AHS+04}.
The matrix elements due to the magnetic dipole and electric quadrupole
corrections are presented in \cite{AHS+04}.

\paragraph{Magnetocrystalline anisotropy energy.}

The internal energy of a ferromagnetic material depends on the direction of
spontaneous magnetization. The magnetocrystalline anisotropy energy (MAE),
which possesses the crystal symmetry of a material, is a part of this energy.
The MAE is an important property that describes the tendency of the magnetization to
align along specific spatial directions rather than to randomly
fluctuate over time. The MAE determines the stability of
magnetization in bulk as well as nanoparticle systems. Extensive
studies on ferromagnetic bulk materials and thin films have
highlighted the MAE dependence on crystal symmetry and atomic
composition. While the exchange interaction among electron spins is
purely isotropic, the orbital magnetization connects the spin magnetization
to the atomic structure of a magnetic material via the spin-orbit interaction,
 giving rise to the magnetic anisotropy \cite{HLO+07}. The calculation of the magnetocrystalline
anisotropy energy has been a long-standing problem. A first theory of
the MAE in Fe and Ni was formulated by Brooks \cite{Brooks40} and
Fletcher \cite{Flet54}, who emphasized that the energy band picture in
which the effect of spin-orbit (SO) coupling is taken into account in
a perturbative way could provide a coupling of the magnetization
orientation to the crystallographic axes of approximately the right
order of magnitude. In this pioneering work, the band structure was
oversimplified to three empirical bands \cite{Brooks40,Flet54}. Recent
investigations \cite{DKS90,SSG+91,TJE+95,HPO+98,RDJ+99} elaborated the
MAE problem using {\it ab initio} calculated energy bands obtained
within the local-spin density approximation to the density functional
theory. Although it is beyond doubt that LSDA energy bands are
superior to empirical bands, it turned out that the calculation of the MAE
from first principles poses a great computational challenge. The prime
obstacle is the smallness of the MAE, which is of only a few meV/atom, a value
that ought to result from the difference of two total energies for
different magnetization directions, which are both of the order of
10$^4$~eV/atom. Due to this numerical problem, it remained
at first unclear if the LSDA could at all describe the MAE correctly,
since the wrong easy axis was obtained for hcp Co and fcc
Ni \cite{DKS90}. Recent contributions aimed consequently at improving
the numerical techniques \cite{TJE+95,WWF93}, with the result that the
correct easy axis was obtained for hcp Co, but not for fcc
Ni \cite{TJE+95}. Halilov  et al. \cite{HPO+98} reported an {\it
  ab initio} investigation of the magnetocrystalline anisotropy
energy in bcc Fe and fcc Co and Ni. They introduced a spin-orbit
scaling technique that yielded the correct easy axis for Fe and Co,
but a vanishing MAE for Ni.

For the material exhibiting uniaxial anisotropy, such as a hexagonal crystal, the MAE can be
expressed as \cite{book:SW59}
\begin{eqnarray}
E(\theta) = K_1 \sin^2 \theta + K_2 \sin^4 \theta + K_3' \sin^6 \theta  %\nonumber \\
+ K_3 \sin^2 \theta \cos\left[6(\phi + \psi)\right] + \cdots, %\label{3}
\end{eqnarray}
where $K_i$ is the anisotropy constant of the $i$th order, $\theta$ and $\phi$
are the polar angles of the Cartesian coordinate system where the $c$ axis
coincides with the $z$ axis (the Cartesian coordinate system is chosen so
that the $x$ axis is rotated through 90$^{\circ}$ from the $a$ hexagonal axis)
and $\psi$ is a phase angle.

Both the magnetic dipole interaction and the SO coupling give rise to the MAE, the
former contributing only to the first-order constant $K_1$. Here, we deal with
the MAE caused only by the SO interaction. Both the magneto-optical effects and
the MAE have a common origin in the SO coupling and exchange splitting. Thus, a
close connection between the two phenomena seems plausible.

In this paper, the MAE is defined as the difference between two
self-consistently calculated fully relativistic total energies for two
different magnetic field directions, $E(\theta)$--$E_{\langle 001\rangle}$.

\paragraph{Crystal structure.}

\begin{figure}[ht]
\hspace{1cm}
\includegraphics[width=0.35\textwidth]{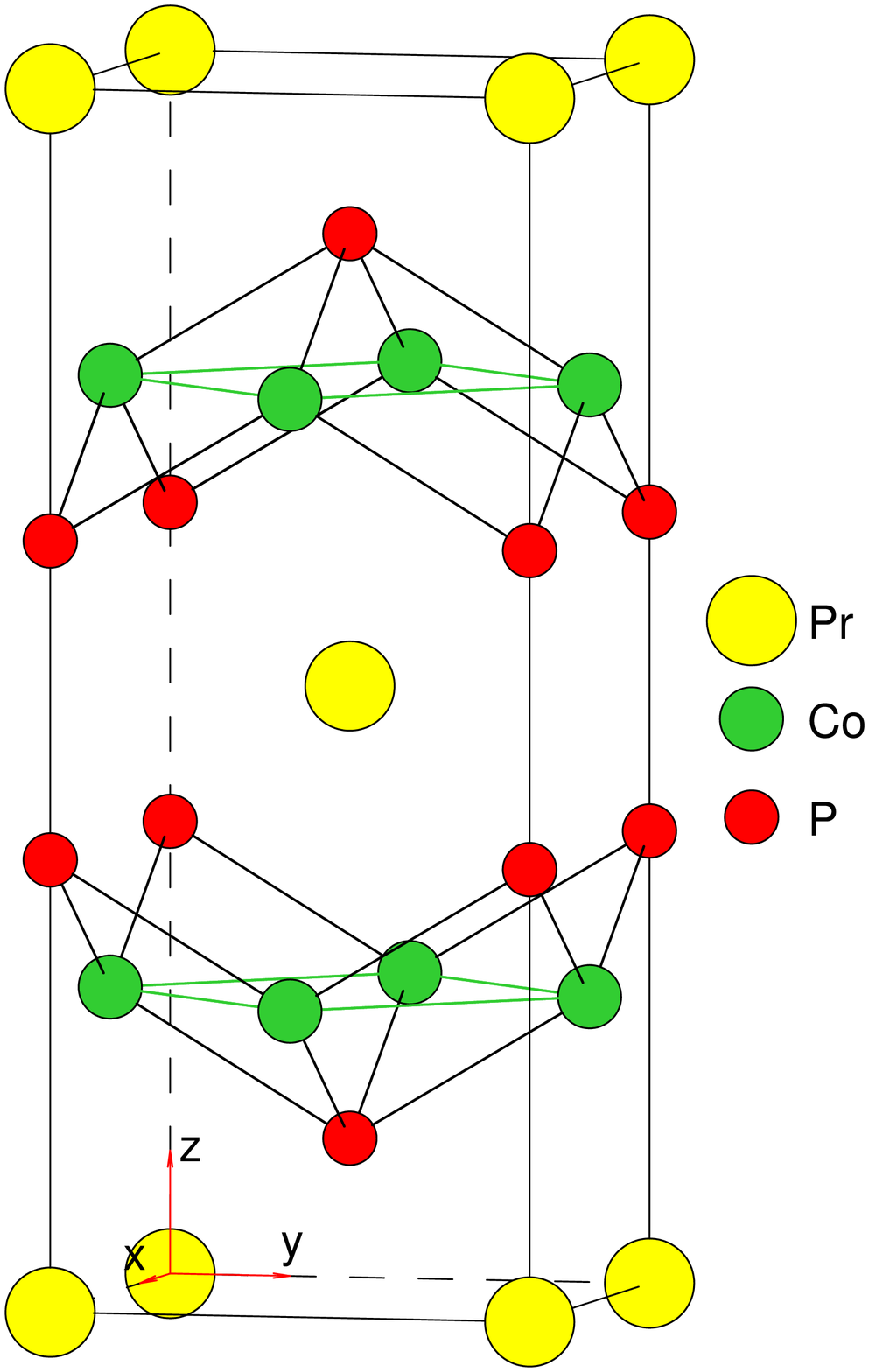}%
\hfill%
\includegraphics[width=0.35\textwidth]{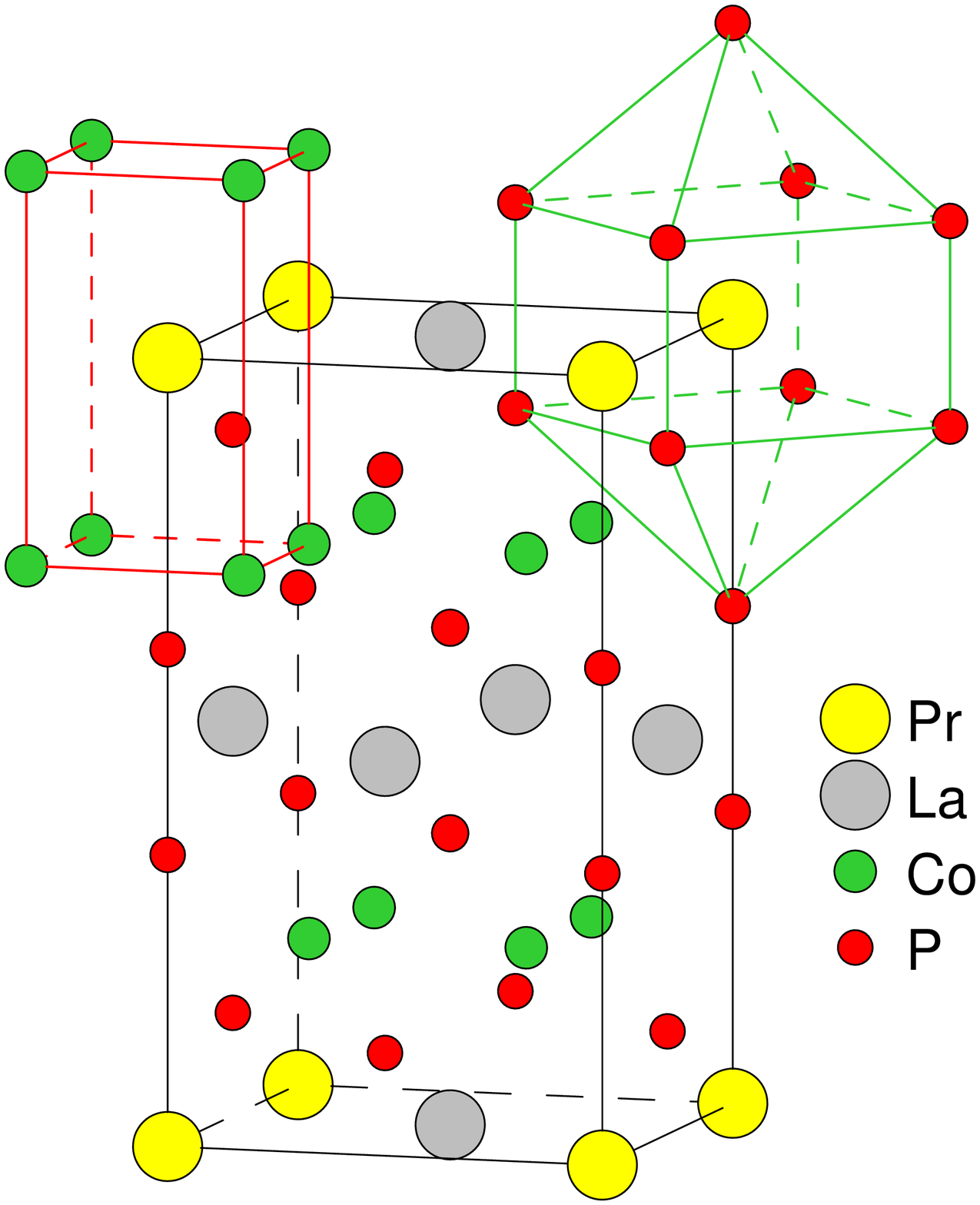}%
\hspace{1cm}
\\%
\parbox[t]{0.48\textwidth}{%
\caption{%
(Color online) Schematic representation of the
  PrCo$_2$P$_2$ structure.
}%
\label{struc}%
}%
\hfill%
\parbox[t]{0.48\textwidth}{%
\caption{%
(Color online) Schematic representation of the
  La$_{1-x}$Pr$_x$Co$_2$P$_2$ ($x=0.25$) structure.
}%
\label{struc_La_Pr}%
}%
\end{figure}
LaCo$_2$P$_2$ and PrCo$_2$P$_2$ belong to the ThCr$_2$Si$_2$ body-centered
tetragonal structure type (figure~\ref{struc}). The space group
is I4/mmm (No.~139) with Pr (La) at the 2$a$ positions (0, 0, 0), Co
at the 4$a$ positions (0, $\frac{1}{2}$, $\frac{1}{4}$) and P at the
4$e$ positions (0, 0, $z$); $z=0.3697$ and 0.3568 in PrCo$_2$P$_2$ and
LaCo$_2$P$_2$, respectively \cite{RBJ+93,EEB+94}. In this structure,
the transition metal atoms (Co) form planar square nets, and the nonmetal atoms
(P) cap the centers of the squares above and below the planes in a
chessboard-like fashion. The resulting [Co$_2$P$_2$] layers are
separated by layers of rare-earth metal cations (Pr or La). The Pr
atom in PrCo$_2$P$_2$ has 8 P nearest neighbors at the distance of
3.0329~\AA\ and 8 Co atoms at the 3.1094~\AA\ distance, the shortest
P--P distance is equal to 2.5247~\AA. In LaCo$_2$P$_2$, the La atom has
8 P nearest neighbors at the distance of 3.1265~\AA\ and 8 Co atoms
at the 3.3551~\AA\ distance, the P--P distance in LaCo$_2$P$_2$ is
much larger and is equal to 3.1621~\AA.

The calculations of the energy band structure of La$_{1-x}$Pr$_x$Co$_2$P$_2$
($x=0, 0.25$, and 1) compounds were performed for $a \times a \times 2c$
supercells of the tetragonal structure with space group of P4/mmm (No. 123).
The structure refinement parameters of La$_{1-x}$Pr$_x$Co$_2$P$_2$ compounds
are presented in \cite{KTZ+10}. The crystal structure of
La$_{0.75}$Pr$_{0.25}$Co$_2$P$_2$ is presented in figure~\ref{struc_La_Pr}. The Pr
atom in this compound has 8 P nearest neighbors at the distance of 3.0329
\AA\ and 8 Co atoms at the 3.1094~\AA\ distance. The Co atoms are surrounded by 4 P atoms at
the distance of 2.2688~\AA\ and 4 Co atoms at the distance of 2.7577~\AA, 4 La atoms are at the 3.1094~\AA\ distance from Co.

\paragraph{Calculation details.}

The calculations presented in this work were performed using the
spin-polarized fully relativistic linear muffin-tin orbital (SPR LMTO)
method \cite{And75,NKA+83,APS+95} for the experimentally observed
lattice constants: $a=3.6$~\AA, $c=9.688$~\AA\ for PrCo$_2$P$_2$
\cite{RBJ+93}, $a=3.8145$~\AA, $c=11.041$~\AA\ for LaCo$_2$P$_2$
\cite{EEB+94}, and $a=3.8260$~\AA, $c=10.9031$~\AA\ for
La$_{1-x}$Pr$_x$Co$_2$P$_2$ ($x=0.25$) \cite{KTG+13}. The basis
consisted of the Pr (La) $s$, $p$, $d$, and $f$; Co $s$, $p$, and $d$;
P $s$, $p$, and $d$ LMTO's. The {\bf k-}space integrations were
performed with the improved tetrahedron method \cite{BJA94}, and the
self-consistent charge density was obtained with 1063 irreducible {\bf
  k-}points.  To attain good convergence in total energy, a large
number of {\bf k} points should be used in calculations. To
resolve the difference in total energies and to investigate the
magnetocrystalline anisotropy, we used 13824 {\bf k} points in the
irreducible part of the Brillouin zone, which corresponds to 82944
tetrahedra in the full zone.

The X-ray absorption and dichroism spectra were calculated taking into
account the exchange splitting of core levels. We also take into
account the core-hole effect in the final state using the supercell
approximation. The similar approximation has been used by several
authors \cite{BAB96,SE98}.
The finite lifetime of a core hole was accounted for by folding the spectra
with a Lorentzian. The widths of core level spectra $\Gamma_{L_2}=1.13$~eV and
$\Gamma_{L_3}=0.47$~eV for Co, and $\Gamma_{M_{4,5}}=0.75$~eV for Pr were taken
from \cite{CaPa01}. The finite apparative resolution of a
spectrometer was accounted for by a Gaussian with the width of 0.6~eV.

It is well known that the LSDA fails to describe the electronic
structure and properties of the systems in which the interaction among
the electrons is strong. In recent years, more advanced methods of
electronic structure determination such as the LSDA plus self-interaction
corrections (SIC--LSDA) \cite{PZ81,AHS+04}, the LSDA+$U$ \cite{AZA91}
method, the GW approximation \cite{Hed65} and the dynamical mean-field
theory (DMFT) \cite{MV89,PJF95,GKK+96} have sought to remedy this
problem and have met considerable success. The LSDA+$U$ method is
the simplest among them and most frequently used in the literature.
We used the ``relativistic'' generalization of the LSDA+$U$ method which
takes into account the spin-orbit coupling so that the occupation matrix
of localized electrons becomes non-diagonal in spin indexes. This
method is described in detail in our previous paper \cite{YAF03}
including the procedure to calculate the screened Coulomb $U$ and
exchange $J$ integrals, as well as the Slater integrals $F^2$, $F^4$,
and $F^6$.

The screened Coulomb $U$ and exchange $J$ integrals enter the LSDA+$U$ energy
functional as external parameters and should be determined independently. The
value of $U$ can be estimated from the photo-emission spectroscopy (PES) and
X-ray Bremsstrahlung Isochromat Spectroscopy (BIS) experiments. Because of the
difficulties with unambiguous determination of $U$, it can be considered as a
parameter of the model. Then, its value can be adjusted to achieve the best
agreement of the results of LSDA+$U$ calculations with PES or optical
spectra \cite{BABH00}. While the use of an adjustable parameter is generally
considered an anathema among first principles practitioners, the LSDA+$U$
approach does offer a plausible and practical method to treat approximately
strongly correlated orbitals in solids.
The Hubbard $U$ and exchange parameter
$J$ can be determined from supercell LSDA calculations using the Slater's
transition state technique \cite{AG91,SDA94} or from constrained LSDA
calculations \cite{DBZ+84,SDA94,PEE98}. Cococcioni and Gironcoli
\cite{CoGi05} have also provided an internally consistent, basis-set independent
method based on the linear response approach for the calculation of the
effective interaction parameters in the LSDA+$U$ method. The constrained LSDA
calculations produce $U=6.17$~eV, $J=0.87$~eV for Pr and $U=4.1$~eV,
$J=0.81$~eV for Co in PrCo$_2$P$_2$, as well as $U=6.08$~eV, $J=0.83$~eV for La
and $U=4.0$~eV, $J=0.8$~eV for Co in LaCo$_2$P$_2$. These values of
$U$ and $J$ were used in our calculations presented below.\\
\begin{wrapfigure}{o}{0.5\textwidth}
\vspace{3mm}
\centerline{\includegraphics[width=6cm]{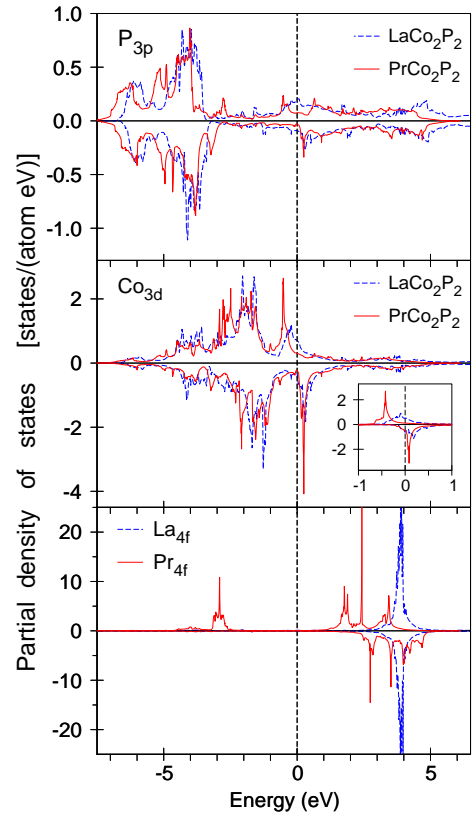}
}
\caption{(Color online) Partial density of states
  of LaCo$_2$P$_2$ and PrCo$_2$P$_2$.
  The insert in the middle panel shows the energy
  distribution of the $xy$ orbitals.}
  \label{PDOS}
  \vspace{-20mm}
\end{wrapfigure}

\section{Results}
\label{sec:results}

\subsection{Electronic structures of LaCo$_2$P$_2$ and PrCo$_2$P$_2$}

Figure~\ref{PDOS} presents the partial density of states (PDOS) for LaCo$_2$P$_2$
and PrCo$_2$P$_2$. The P 3$s$ states are located
mostly between $-13.6$~eV and $-10.0$~eV below the Fermi level (not shown), while the
3$p$ states of P are found between $-6.8$~eV and 6.1~eV in LaCo$_2$P$_2$
and between $-7.5$~eV and 5.8~eV in PrCo$_2$P$_2$. The spin splitting of the P
3$p$ states is quite small. The Co 3$d$ states occupy the energy interval between
$-6.9$~eV and 5.5~eV and hybridize strongly with the P 3$p$ states. The La
4$f$ empty states occupy the $3.2-4.5$~eV energy interval in LaCo$_2$P$_2$.
The Pr 4$f$ empty states are from 1.5~eV to 4.8~eV above the Fermi level. Two Pr
4$f$ spin-up electrons occupy a small energy interval around $-3$~eV.

The crystal field at the Co site ($D_{2d}$ point symmetry) in both compounds
causes the splitting of Co 3$d$ orbitals into three singlets $a_1$ ($\dzz$),
$b_1$ ($\dxy$), $b_2$ ($\dxxyy$) and a doublet $e$ ($\dyz$ and $\dxz$). A two
peak structure of the minority- and majority-spin Co $d$ states is found
in the close vicinity of the Fermi energy.
We found that the two peaks (occupied in the majority-spin channel at $-0.5$~eV and empty in the minority-spin channel at 0.3~eV) are of the $xy$ character. These peaks are rather
broad in LaCo$_2$P$_2$ and very narrow and intensive in PrCo$_2$P$_2$. Such
a difference can be explained by different nearest-neighbors interatomic
Co$-$Co and Co$-$P distances. The Co$-$Co distance is equal to 2.697~\AA\ and
2.757~\AA\ in LaCo$_2$P$_2$ and PrCo$_2$P$_2$, respectively. Also, the Co$-$P
distance is larger in LaCo$_2$P$_2$ than in PrCo$_2$P$_2$ by 0.027~\AA.

\begin{figure}[ht]
\hspace{2cm}
\includegraphics[width=0.25\textwidth]{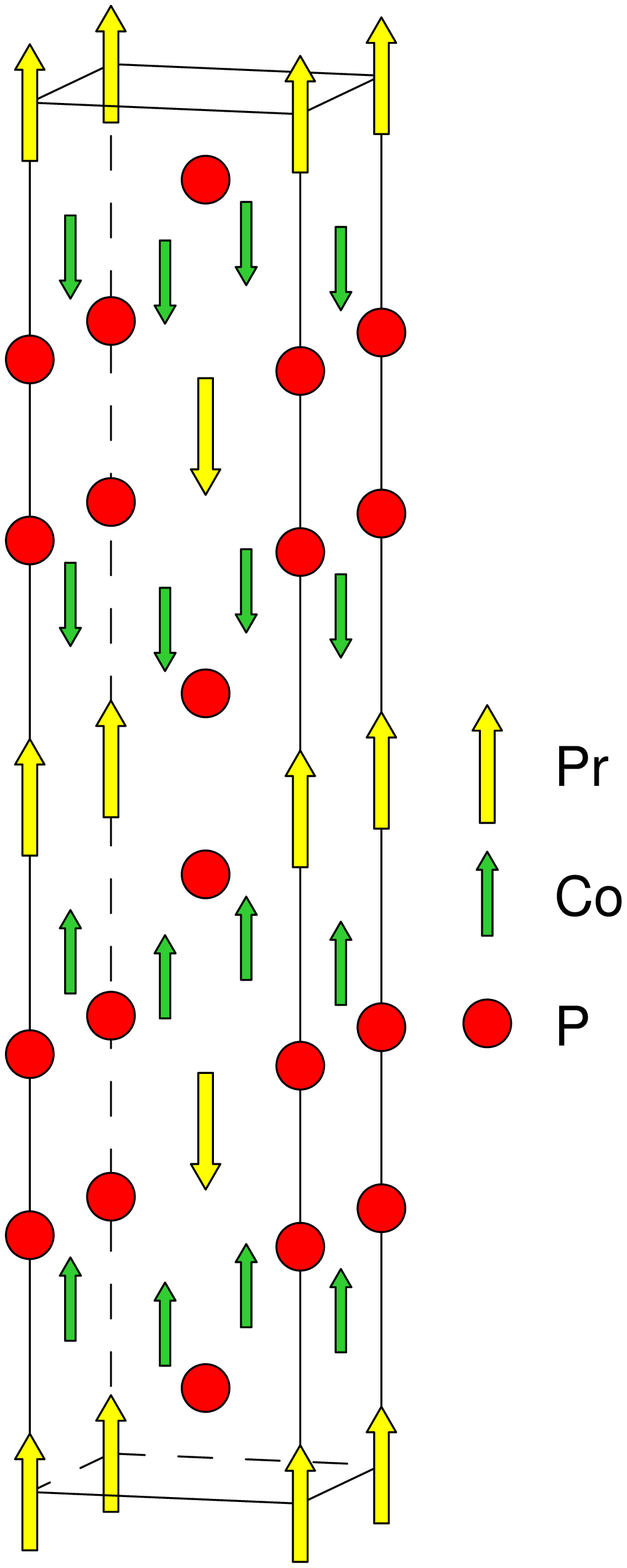}%
\hfill%
\includegraphics[width=0.25\textwidth]{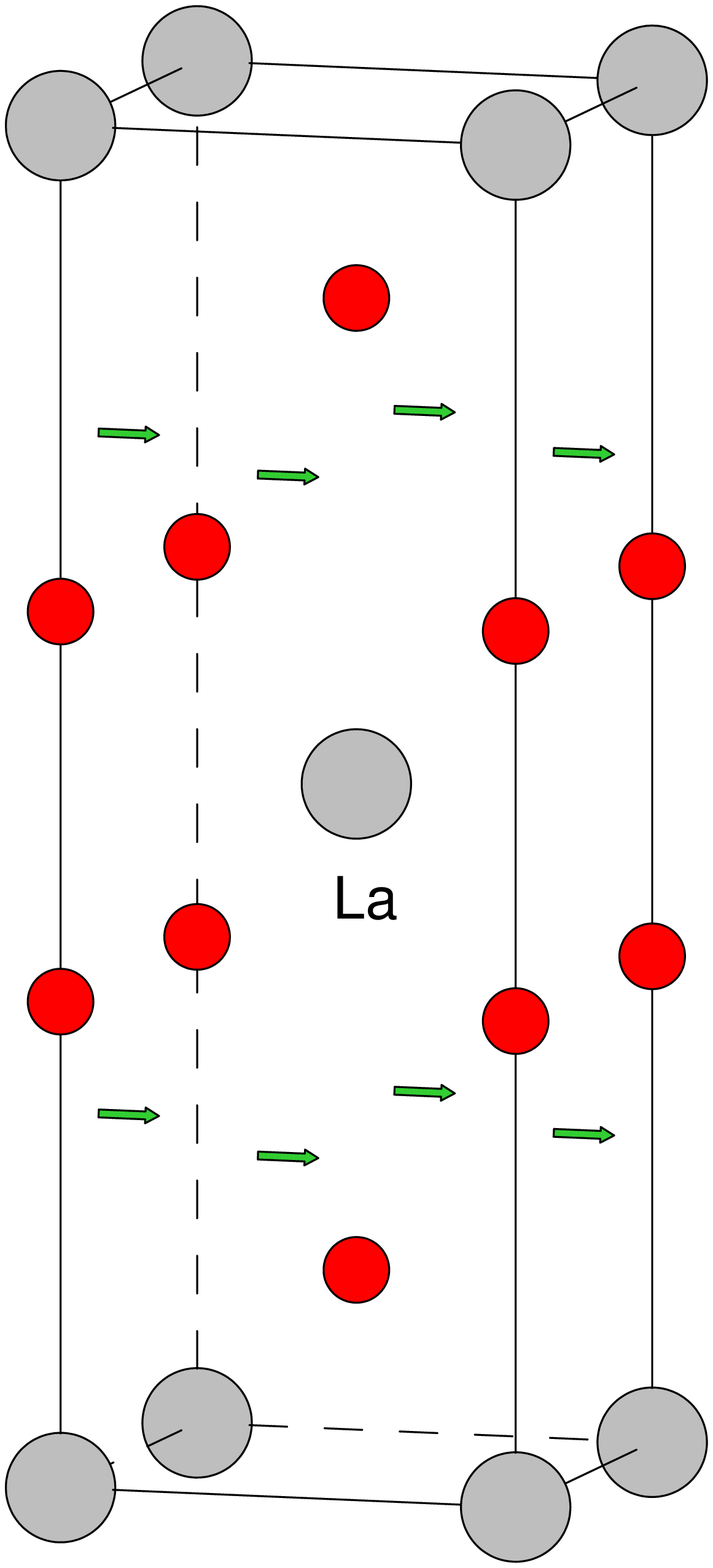}%
\hspace{2cm}
\\%
\parbox[t]{0.48\textwidth}{%
\centerline{(a)}%
}%
\hfill%
\parbox[t]{0.48\textwidth}{%
\centerline{(b)}%
}%
\caption{(Color online) Schematic representation of the
  magnetic ordering in PrCo$_2$P$_2$ (left-hand column) and LaCo$_2$P$_2$ (right-hand
  column).}
\label{struc_M}
\end{figure}
Our self-consistent calculations reveal a ferromagnetic arrangement of
magnetic moments in PrCo$_2$P$_2$. The spin magnetic moments at the Pr and Co
sites are aligned along the $c$ axis with the antiparallel magnetic coupling
between the neighboring planes resulting in antiferromagnetism (left-hand column of
figure~\ref{struc_M}). The similar magnetic arrangement was observed experimentally
by Reehuis  et al. \cite{RBJ+93}. The theoretical calculations give a
ferromagnetic arrangement of magnetic moments in LaCo$_2$P$_2$, the Co moments
are aligned in-plane and parallel to the moments in the other layers (right-hand
column of figure~\ref{struc_M}). Figure~\ref{MAE} (upper panel) shows the MAE
as a function of the polar angle $\theta$. The minimum of
the total energy corresponds to the magnetic configuration with the Co moments
aligned in-plane in agreement with the experimental observation
\cite{EEB+94}. The theory produces quite a small value of MAE of around
0.28 meV per formula unit in LaCo$_2$P$_2$. The lower panel of figure~\ref{MAE} shows the anisotropy
of spin magnetic moments at the Co site in LaCo$_2$P$_2$ as a function of the
polar angle $\theta$.

The spin $m_s$ and orbital $m_l$ magnetic moments at the Co site in
PrCo$_2$P$_2$ are larger than in LaCo$_2$P$_2$. Our band structure
calculations yield the magnetic moments for the Co atoms $m_s$ = 0.825 {\mb},
$m_l$ = 0.120 {\mb} in PrCo$_2$P$_2$ and $m_s$ = 0.665 {\mb}, $m_l$ = 0.066
{\mb} in LaCo$_2$P$_2$. The induced spin magnetic moments at the P site are of
0.011 {\mb} and 0.016 {\mb} for the PrCo$_2$P$_2$ and LaCo$_2$P$_2$,
respectively. The orbital moments at the P sites are small in both
compounds ($m_l^{\rm P}=0.005$~{\mb}). The orbital magnetic moment at the La
site is extremely small $m_l^{\rm La}=-0.002$~{\mb}, however, we found
quite a large orbital moment at the Pr site in PrCo$_2$P$_2$ ($m_l^{\rm
  Pr}=0.509$~{\mb}).

\subsection{Electronic structure, X-ray absorption and XMCD spectra in
La$_{1-x}$Pr$_x$Co$_2$P$_2$ compounds }

\begin{figure}[ht]
\vspace{2ex}
\hspace{5mm}
\includegraphics[width=0.4\textwidth]{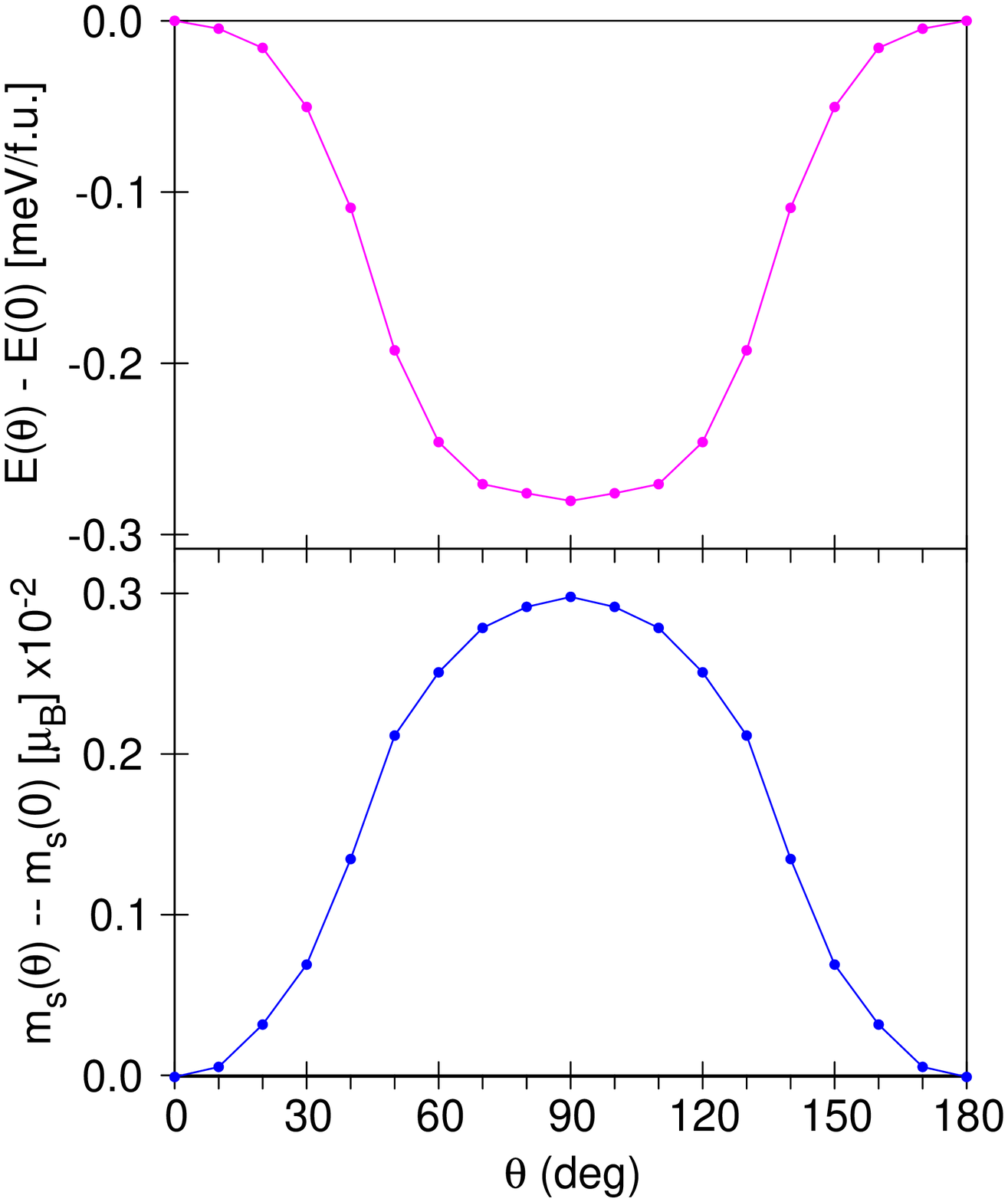}%
\hfill%
\includegraphics[width=0.4\textwidth]{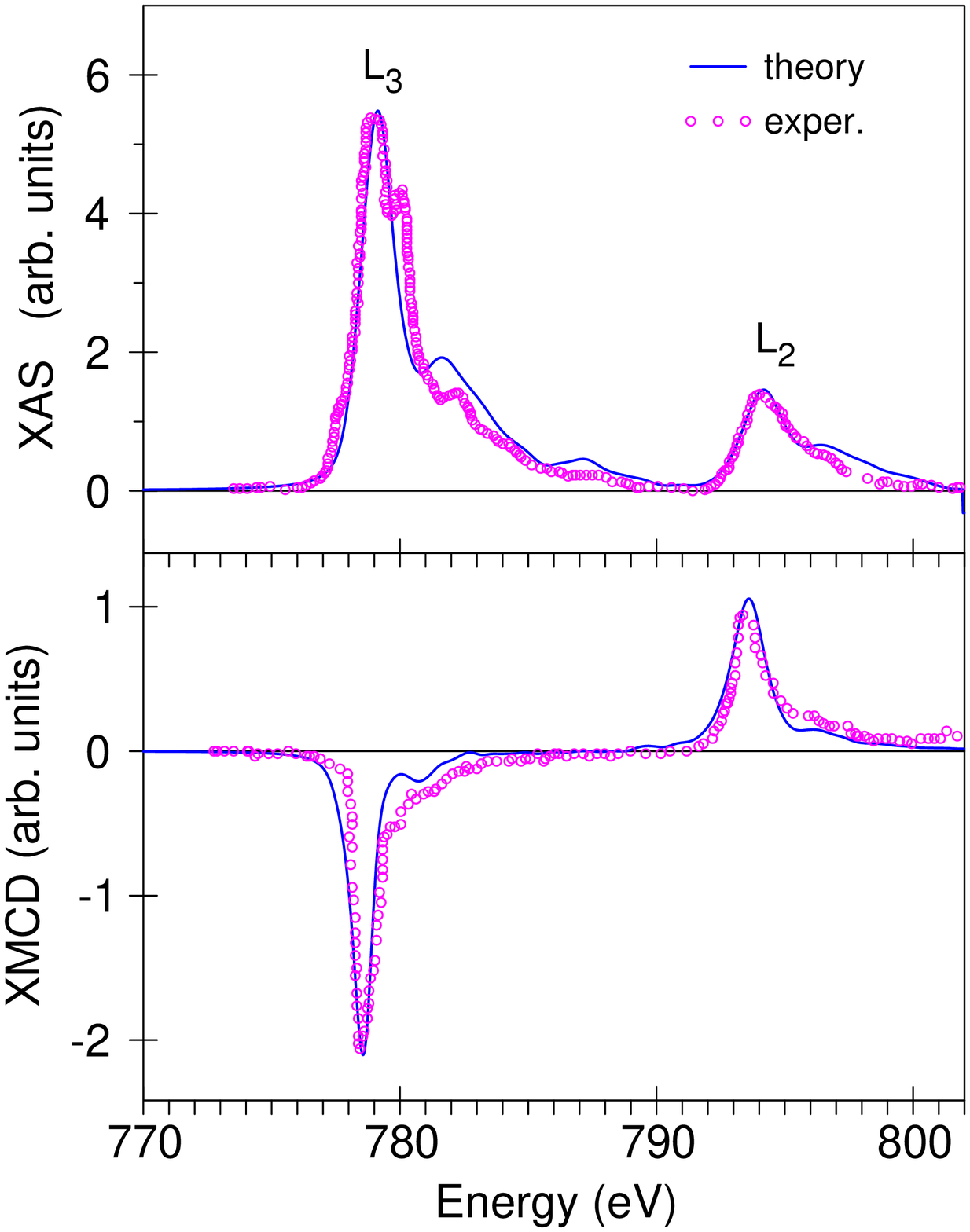}%
\hspace{5mm}
\\%
\parbox[t]{0.48\textwidth}{%
\caption{%
(Color online) MAE (upper panel) and
  anisotropy of spin magnetic moments at the Co site (lower panel) in LaCo$_2$P$_2$
  as functions of the polar angle $\theta$.
}%
\label{MAE}%
}%
\hfill%
\parbox[t]{0.48\textwidth}{%
\caption{%
(Color online) Partial density of states
  in La$_{1-x}$Pr$_x$Co$_2$P$_2$ for $x=0.25$.
}%
\label{DOS}%
}%
\vspace{2ex}
\end{figure}

Figure~\ref{DOS} presents the partial density of states in
La$_{1-x}$Pr$_x$Co$_2$P$_2$ for $x=0.25$. The energy position and shape of
PDOS slightly differ from the corresponding PDOS of LaCo$_2$P$_2$ and
PrCo$_2$P$_2$. The P 3$p$ PDOS has more pronounced peaks near the Fermi level
due to the larger P 3$p$ $-$ Co 3$d$ hybridization. The positions of the empty La and Pr
4$f$ states are shifted closer to the Fermi level in comparison with the reference
compounds. Also, the shape of the energy distribution of Pr 4$f$ states
slightly differ from the corresponding 4$f$ PDOS in PrCo$_2$P$_2$ and
LaCo$_2$P$_2$. Such differences might be explained by different lattice
constants and different numbers and distances of the nearest neighbors.
%%%%%%%%%%%%%%%%%%%%%%%%%

The spin magnetic moments at the Pr and Co sites (1.955 {\mb} and 0.815 {\mb},
respectively) are aligned along the $c$ axis in
La$_{0.75}$Pr$_{0.25}$Co$_2$P$_2$ with antiparallel magnetic moments at the La
sites ($m_s = -0.002$~{\mb} and $-0.007$~{\mb} for two nonequivalent La
sites). The orbital magnetic moments are equal to 0.054 {\mb}, 0.109 {\mb},
and 0.020 {\mb} for Pr, Co, and La sites, respectively. The orbital moments at
the P sites are small ($m_l^{\rm P}=0.001$~{\mb}).

\begin{figure}[ht]
\hspace{3mm}
\includegraphics[width=0.4\textwidth]{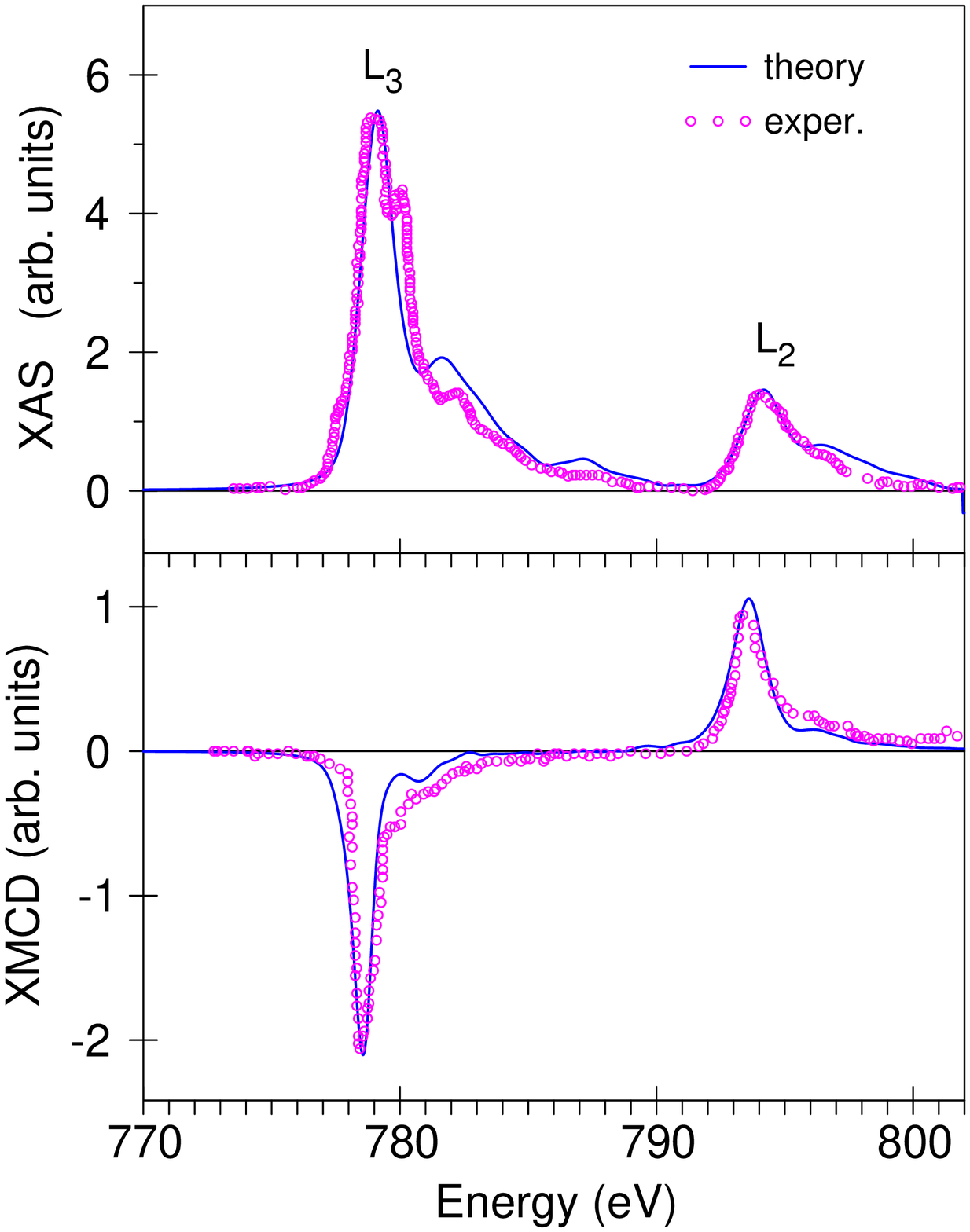}%
\hfill%
\includegraphics[width=0.4\textwidth]{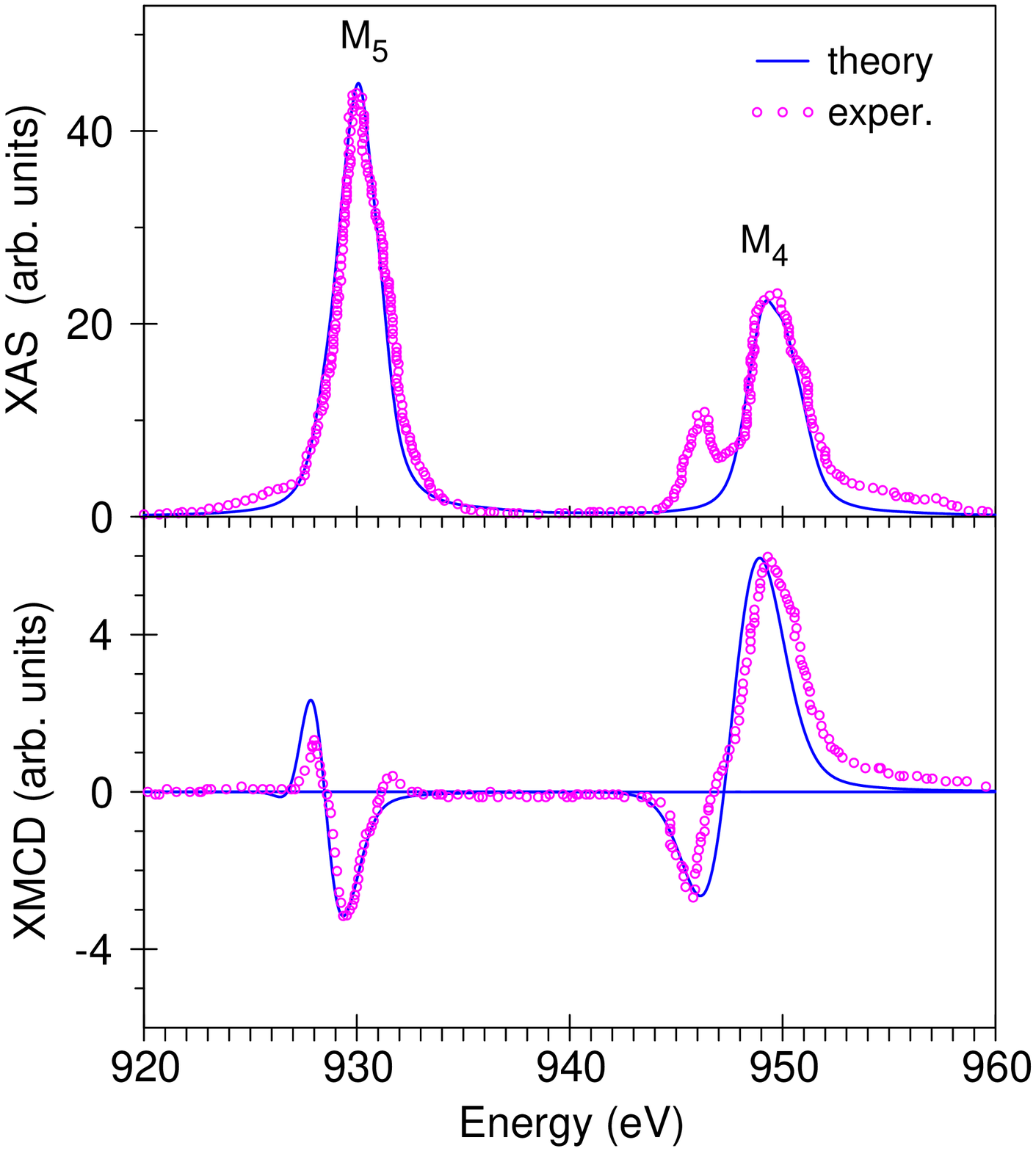}%
\hspace{8mm}
\\%
\parbox[t]{0.48\textwidth}{%
\caption{%
(Color online) Experimental and theoretically
  calculated X-ray absorption (upper panel) and XMCD (lower panel) spectra
  of La$_{0.75}$Pr$_{0.25}$Co$_2$P$_2$ at the Co $L_{2,3}$ edges.
}%
\label{xmcd_Co}%
}%
\hfill%
\parbox[t]{0.48\textwidth}{%
\caption{%
(Color online) Experimental and theoretically
  calculated X-ray absorption (upper panel) and XMCD (lower panel) spectra
  of La$_{0.75}$Pr$_{0.25}$Co$_2$P$_2$ at the Pr $M_{4,5}$ edges.
}%
\label{xmcd_Pr}%
}%
\end{figure}

Figure \ref{xmcd_Co} (upper panel) shows the X-ray absorption spectra
at the Co $L_{2,3}$ edges in La$_{0.75}$Pr$_{0.25}$Co$_2$P$_2$ measured
at 5 K \cite{KTG+13} with a 1 T magnetic field applied along the $c$ axis
compared with the theoretically calculated ones in the
LSDA+$U$ approximation. The Co $L_3$ X-ray absorption spectrum possesses four
major fine structures: a major peak at 779~eV and three high energy shoulders at
780.5~eV, 782.5~eV and 787.5~eV. The theory describes reasonably well the
energy position and relative intensity of all the fine structures except the
shoulder at 780.5~eV. Due to the electric dipole selection rules ($\Delta
l=\pm 1$; $\Delta j =0,\, \pm 1$), the major contribution to the absorption at
the $L_3$ edge stems from the transitions 2$p_{3/2}$ $\to$ 5$d_{5/2}$, with a
weaker contribution from the 2$p_{3/2}$ $\to$ 5$d_{3/2}$ transitions. For the
latter case, the corresponding 2$p_{3/2}$ $\to$ 5$d_{3/2}$ radial matrix
elements are only slightly smaller than for the 2$p_{3/2}$ $\to$ 5$d_{5/2}$
transitions. The angular matrix elements, however, strongly suppress the
2$p_{3/2}$ $\to$ 5$d_{3/2}$ contribution. Therefore, the contribution to the
XAS spectrum at the $L_3$ edge from the transitions with $\Delta j=0$ is one
order of magnitude smaller than the transitions with $\Delta j=1$
\cite{book:AHY04}.

The lower panel of figure~\ref{xmcd_Co} presents the XMCD experimental spectra
of La$_{0.75}$Pr$_{0.25}$Co$_2$P$_2$ at the Co $L_{2,3}$ edges and
the theoretically calculated ones. The LSDA+$U$ calculations describe
reasonably well all the features of the experimental XMCD spectra.

A study of the 4$f$ electron shell in rare earth compounds is usually
performed by tuning the energy of X-ray close to the $M_{4,5}$ edges of
rare-earth where the electronic transitions between 3$d_{3/2,5/2}$ and
4$f_{5/2,7/2}$ states are involved. Figure~\ref{xmcd_Pr} shows the XAS and XMCD
spectra at the Pr $M_{4,5}$ edges in La$_{0.75}$Pr$_{0.25}$Co$_2$P$_2$
measured at 5 K \cite{KTG+13} with a 1 T magnetic field applied along the $c$
axis compared with the theoretically calculated ones in the LSDA+$U$
approximation. The Dirac-Hartree-Fock-Slater single-particle
approximation used in this work to calculate the core states is not capable of
producing a correct energy position of the spectra (because the self-interaction correction, different kinds of relaxation processes
and other many-particle effects were not taken into
account). Therefore, we used the experimentally
measured positions of the spectra. The theoretically calculated XAS spectra
have a rather simple line shape composed of two white line peaks at the $M_5$
and $M_4$ edges. The theory reproduces the shape of the $M_5$ XAS spectrum very
well. However, the experimentally measured $M_4$ spectrum has a well pronounced
fine structure at the low energy part that is not reproduced by the theory. We
should mention here that a major shortcoming in the band structure approximation
is that the multiplet structure is not included. For the Co $L_{2,3}$
edges, this is not the major problem. However, for the Pr $M_{4,5}$ edges,
the core-valence electrostatic interactions can significantly effect the
line shape of the XAS and XMCD spectra. The fine structure on the low energy
side of Pr $M_4$ XASs is believed to be due to the multiplet structure, which
is not included in our calculations. A theoretical method that
 consistently includes both the band structure and the atomic-like multiplet
structure of rare earth metals and compounds is highly desired. However, we
should mention that it is still not clear why the multiplet structure is
pronounced only at the Pr $M_4$ edge and is not seen at the $M_5$ edge.
%%%%%%%%%%%%%%%%%%%%%%%%%

Figure \ref{xmcd_Pr} (lower panel) shows the calculated Pr $M_{4,5}$ XMCD
spectra in the LSDA+$U$ approximation for La$_{0.75}$Pr$_{0.25}$Co$_2$P$_2$
together with the corresponding experimental data \cite{KTG+13}. The
experimentally measured dichroism is rather large. The XMCD spectra at the
$M_5$ and $M_4$ edges have a two-peak structure. The dichroism is positive at a
lower energy and negative at a higher energy at the Pr $M_5$ edge, while the Pr $M_4$
XMCD spectrum has a negative minimum at a lower energy and a positive maximum at a higher
energy. The LSDA+$U$ calculations quite well describe all the features of
the experimental XMCD spectra at the $M_{4,5}$ edges. The theory also
correctly reproduces the relative intensities of the XAS and XMCD spectra at
the $M_5$ and $M_4$ edges, namely, the XAS is larger at the $M_5$ edge than at
the $M_4$ edge, while the XMCD spectra show the opposite behavior with larger dichroism
at the $M_4$ edge in comparison with the $M_5$ edge.

When a 3$d$ core-electron is photo-excited to an unoccupied 4$f$ state, the distribution
of the charge changes to account for the created hole. We investigated this core-hole effect
in the final state using the supercell approximation. We found that the final-state interaction has little
effect on the shape of the XAS and XMCD spectra at the Pr $M_{4,5}$ edges.
We also investigated the effect of the electric quadrupole $E_2$ and magnetic
dipole $M_1$ transitions. We found that the $M_1$ transitions are extremely
small in comparison with the $E_2$ transitions and can be neglected. The $E_2$
transitions are much weaker than the electric dipole transitions $E_1$. They are almost
invisible in the XAS and have a very small effect on the XMCD spectra at the Pr
$M_{4,5}$ edges.

\section{Summary}
\label{sec:summ}

The electronic structure and magnetic ordering in La$_{1-x}$Pr$_x$Co$_2$P$_2$
($x=0, 0.25$, and 1) phosphides have been studied theoretically using the
fully relativistic spin-polarized Dirac LMTO band-structure method.  The
self-consistent calculations reveal a ferromagnetic arrangement of magnetic
moments in PrCo$_2$P$_2$. The spin magnetic moments at the Pr and Co sites are
aligned along the $c$ axis with the antiparallel magnetic coupling between
the neighboring planes resulting in antiferromagnetism. LaCo$_2$P$_2$ possesses a
ferromagnetic arrangement of magnetic moments where the Co moments are aligned
in-plane and parallel to the moments in the other layers. The theory gives
quite a small value of MAE in LaCo$_2$P$_2$ (around 0.28 meV per formula unit).

We have studied the X-ray magnetic circular dichroism at the Co $L_{2,3}$ and Pr
$M_{4,5}$ edges in \linebreak La$_{0.75}$Pr$_{0.25}$Co$_2$P$_2$. The calculations show
good agreement with the experimental measurements. The core-hole effect was
found to be very small on the shape of the XAS and XMCD spectra at the Pr
$M_{4,5}$ edges. We found that the magnetic dipole $M_1$ transitions are
extremely small in comparison with the electric quadrupole $E_2$ transitions
and can be neglected. The $E_2$ transitions were found to be weaker than the electric
dipole transitions $E_1$. They are almost invisible in the XAS and have a very small
effect on the XMCD spectra at the Pr $M_{4,5}$ edges.

%\bibliography{./jprb,./book,./LaPrCoP}
%\bibliography{./jprb,./book,./Yb_Ag_Cu,./LaPrCoP,./MnBi}

\newcommand{\noopsort}[1]{} \newcommand{\printfirst}[2]{#1}
  \newcommand{\singleletter}[1]{#1} \newcommand{\switchargs}[2]{#2#1}

\ukrainianpart \label{ukr}
\title{Електронна структура, магнітне впорядкування та рентгенівський магнітний циркулярний дихроїзм у фосфідах La$_{1-x}$Pr$_x$Co$_2$P$_2$}
\author{Л.В. Бекеньов, С.В. Мокляк, В.М. Антонов}

\address{Інститут металофізики ім. Г.В.Курдюмова НАН України, бульв. Вернадського 36, 03142, Київ}
\makeukrtitle
\begin{abstract}
\tolerance=3000%
На основі зонних розрахунків повністю релятивістським спін-поляризованим лінійним методом МТ-орбіталей (ЛМТО) теоретично вивчені електронна структура і магнітне впорядкування у фосфідах La$_{1-x}$Pr$_x$Co$_2$P$_2$ ($x=0, 0.25$, та 1). В рамках методу LSDA+$U$ теоретично досліджені рентгенівські спектри поглинання та спектри рентгенівського циркулярного дихроїзму на краях поглинання Co$L_{2,3}$ і Pr$M_{4,5}$. Вивчено вплив остовної дірки в кінцевому стані, а також електричних квадрупольних $E_2$ і магнітних дипольних $M_1$ переходів. Отримано добре узгодження з експериментальними результатами.
\keywords сильно скорельовані системи, зонна структура, магнітні моменти, рентгенівський магнітний циркулярний дихроїзм
\end{abstract}
\end{document}